\input harvmac
\input epsf.tex
\noblackbox
\overfullrule=0pt
\def\Title#1#2{\rightline{#1}\ifx\answ\bigans\nopagenumbers\pageno0\vskip1in
\else\pageno1\vskip.8in\fi \centerline{\titlefont #2}\vskip .5in}

%
%
%

\def\[{\left [}
\def\]{\right ]}
\def\({\left (}
\def\){\right )}

\def\p{\partial}

\font\cmss=cmss10 \font\cmsss=cmss10 at 7pt
\def\IZ{\relax\ifmmode\mathchoice
   {\hbox{\cmss Z\kern-.4em Z}}{\hbox{\cmss Z\kern-.4em Z}}
   {\lower.9pt\hbox{\cmsss Z\kern-.4em Z}}
   {\lower1.2pt\hbox{\cmsss Z\kern-.4em Z}}\else{\cmss Z\kern-.4emZ}\fi}

%

\lref\arl{H. Araki and E. Lieb, Comm. Math. Phys. {\bf 18} (1970) 160.}
\lref\dasmathurtwo{S. Das and S. Mathur,
``{\it Interactions Involving D-branes}'',
hep-th/9607149.}

\lref\gk{ S. Gubser and I. Klebanov, 
``{\it Emission of Charged
Particles
from Four- and Five-dimensional Black Holes}'', 
hep-th/9608108.}

\lref\page{ D. Page, Phys. Rev. {\bf D 13} (1976) 198; Phys. Rev. {\bf
D14}
(1976), 3260.}

\lref\unruh{W. Unruh, 
Phys. Rev. {\bf D14} (1976) 3251.}
\lref\cj{ M. Cvetic and D. Youm, hep-th/9512127.}

\lref\tseytlin{A. Tseytlin, hep-th/9601119.}
\lref\tata{S. Dhar, G. Mandal and S. Wadia, ``{\it
Absorption vs. Decay of Black Holes in String Theory and 
T-symmetry}'', hep-th/9605234.}
\lref\gm{G. Horowitz and D. Marolf, hep-th/9605224.}
\lref\mss{J. Maldacena and L. Susskind, ``{\it D-branes and Fat 
Black Holes }'', Nucl. Phys. {\bf B475} (1996) 679,  hep-th/9604042.}
\lref\bl{V. Balasubramanian and F. Larsen, hep-th/9604189.}
\lref\jmn{J. Maldacena,``{\it Statistical Entropy of Near Extremal
Fivebranes}'', hep-th/9605016.}
\lref\hlm{G. Horowitz, D. Lowe and J. Maldacena, ``{\it Statistical
Entropy of Nonextremal Four Dimensional Black Holes and U-duality}'',
Phys. Rev. Lett. {\bf 77} (1996) 430, hep-th/9603195.}
\lref\ms{J. Maldacena and A. Strominger, Phys. Rev.  Lett. {\bf 77}
(1996) 428,  hep-th/9603060.}
\lref\bd{See N. D. Birrel and P.C. Davies,``{\it Quantum Fields in
Curved
Space}'', Cambridge University Press 1982.}
\lref\hms{G. Horowitz, J. Maldacena and A. Strominger,
``{\it Nonextremal Black Hole Microstates and U-duality}'', 
 Phys. Lett. {\bf B383} (1996) 151,
 hep-th/9603109.}
\lref\dbr{J. Polchinski, S. Chaudhuri, and C. Johnson, hep-th/9602052.}
\lref\jp{J. Polchinski, hep-th/9510017.}
\lref\dm{S. Das and S. Mathur, ``{\it
Comparing Decay Rates for Black Holes and D-branes}'',hep-th/9606185;
``{\it Interactions Involving D-branes}'',
hep-th/9607149.}
\lref\ghas{G. Horowitz and A. Strominger,``{\it Counting States of
Near-extremal Black Holes }'', Phys. Rev. Lett. {\bf 77} (1996) 2368,
  hep-th/9602051.}
\lref\ascv{A. Strominger and C. Vafa, ``{\it On the Microscopic 
Origin of the Bekenstein-Hawking Entropy}'', Phys. Lett. {\bf B379} (1996)
99,
 hep-th/9601029.}
\lref\spin{
 J.C. Breckenridge, R.C. Myers, A.W. Peet and  C. Vafa, { \it
D-Branes and Spinning Black Holes}'',  hep-th/9602065.}
\lref\clifford{C. Johnson, R. Khuri and R. Myers, {\it 
Entropy of 4D Extremal Black Holes}'', Phys. Lett. {\bf B378} (1996) 78,
 hep-th/9603061.}

\lref\hrva{P. Horava, Phys. Lett. {\bf B231} (1989) 251.}
\lref\cakl{C. Callan and I. Klebanov, hep-th/9511173.}
\lref\prskll{J. Preskill, P. Schwarz, A. Shapere, S. Trivedi and
F. Wilczek, Mod. Phys. Lett. {\bf A6} (1991) 2353. }
\lref\bhole{G. Horowitz and A. Strominger,
Nucl. Phys. {\bf B360} (1991) 197.}
\lref\bekb{J. Bekenstein, Phys. Rev {\bf D12} (1975) 3077.}
\lref\hawkirr{S. Hawking, Phys. Rev {\bf D13} (1976) 191.}
\lref\stas{A.~Strominger and S.~Trivedi,  Phys.~Rev. {\bf D48}
 (1993) 5778.}
\lref\bek{J. Bekenstein, Lett. Nuov. Cimento {\bf 4} (1972) 737,
Phys. Rev. {\bf D7} (1973) 2333, Phys. Rev. {\bf D9} (1974) 3292.}
\lref\hawkb{S. Hawking, Nature {\bf 248} (1974) 30, Comm. Math. Phys.
{\bf 43} (1975) 199.}
\lref\cm{C. Callan and J. Maldacena, ``{\it The D-brane approach to 
black hole quantum mechanics}'', Nucl. Phys. {\bf B 475 } (1996)
645, hep-th/9602043.}
\lref\hpc{S. Hawking, private communication.}
\lref\bdpss{T. Banks, M. Douglas, J. Polchinski, 
S. Shenker and A. Strominger, in progress.}
\lref\ast{A. Strominger, {\it Statistical Hair on Black Holes''},
hep-th/9606016.  }
\lref\ka{ B. Kol and A. Rajaraman, ``{\it
Fixed Scalars and Suppression of Hawking Evaporation}'',  hep-th/9608126. }

\lref\thooft{G. 't Hooft, 
``{\it The Scattering Matrix Approach for the Quantum Black Hole, 
an Overview}'', gr-qc/9607022.}

\lref\vafainst{ M. Bershadsky, V. Sadov and  C. Vafa, ``{\it D-Strings on
D-Manifolds}'', Nucl. Phys. {\bf B463} (1996) 398,  hep-th/9511222;
C.  Vafa, ``{\it Instantons on D-Branes}'',  Nucl. Phys. {\bf B463} (1996) 435,
hep-th/9512078.}

\lref\becspin{  J. C. Breckenridge, D. A. Lowe, 
R. C. Myers, A. W. Peet, A. Strominger and  C. Vafa, ``{\it
Macroscopic and Microscopic Entropy of Near-Extremal Spinning Black
Holes}'', Phys. Lett. {\bf B381} (1996) 423,  hep-th/9603078.}

\lref\jmas{J. Maldacena and A. Strominger, ``{\it Black Hole Greybody
Factor and D-brane Spectroscopy}'', hep-th/9609026.}

\lref\cgkt{C. Callan, S. Gubser,  I. Klebanov and A. Tseytlin, 
``{\it Absorption
of Fixed Scalars and the D-Brane Approach to Black Holes}'', hep-th/9610172.}

\lref\verlindemoore{R. Dijkgraaf, G, Moore, E. Verlinde and H. Verlinde, 
``{\it Elliptic Genera of Symmetric Products and Second Quantized
Strings}'',
hep-th/9608096.}

\lref\verlindecount{ R. Dijkgraaf, E. Verlinde and H. Verlinde, ``{\it
Counting Dyons in N=4 String Theory}'',
 hep-th/9607026.}

\lref\nocouplings{ B. de Wit, P. Lauwers and A. Van Proeyen, 
``{\it Lagrangians of N=2 Supergravity Matter Systems''}, Nucl. Phys.
{\bf B255} (1985) 569.}

\lref\seibergho{N. Seiberg, ``{\it Naturalness vs. Supersymmetric 
Non-renormalization Theorems}'', Phys. Lett. {\bf B318} (1993) 469,
 hep-th/9309335; ``{\it Power of Holomorphy- Exact Results 
in 4D Supersymmetric
Field Theories}'', PASCOS 1994, 357-369 (QCD161:I69:1994), hep-th/9408013.}

\lref\seibergthree{N. Seiberg and E. Witten, 
``{\it Gauge Dynamics and Compactification to Three Dimensions}'', 
hep-th/9607163.}

\lref\hawkingunitarity{ S. W. Hawking, {\it Breakdown of
Predictability
in Gravitational Collapse}'', Phys. Rev. {\bf D14} (1976) 2460.}


\lref\metricfive{T. Banks, M. Douglas, J. Polchinski, S. Shenker and 
A. Strominger, private comunication.}

\lref\douglas{ M. Douglas, {\it Branes within Branes},
hep-th/9512077.}

\lref\polchinski{
J. Polchinski, ``{\it Monopole Catalysis: The Fermion Rotor System}'',
Nucl. Phys. {\bf B242} (1984) 345; Rubakov, Nucl. Phys. {\bf B203}
(1982)
311; C. Callan, Phys. Rev. {\bf D25} (1982) 2141; {\bf D26} (1982)
2058;
Nucl. Phys. {\bf B212} (1983) 391.}

\lref\jmthesis{J. Maldacena, ``{\it Black Holes in String Theory}'',
Ph.D. thesis, Princeton University 1996, hep-th/9608235.}

\lref\hawkingbck{Reference on the area law.}

\lref\notesondbranes{J. Polchinski, S. Chaudhuri and  C. Johnson,
{\it Notes on D-Branes}, hep-th/9602052.}

\lref\sm{ J. Maldacena and A. Strominger, ``{\it
Statistical Entropy of Four-Dimensional Extremal Black Holes}'',
Phys. Rev. Lett. {\bf 77} (1996) 428, 
 hep-th/9603060.}

\lref\ktmbranes{ I. Klebanov and A. Tseytlin, ``{\it Intersecting M-Branes
as Four Dimensional Black Holes}''
Nucl. Phys. {\bf B475} (1996) 179,
 hep-th/9604166; V. Balasubramanian and F. Larsen, 
``{\it On D-Branes and Black Holes in Four Dimensions}'', hep-th/9604189. }  

\lref\gkfour{ S. Gubser and I. Klebanov, ``{\it 
Four Dimensional Grey Body Factors and the Effective String}'', 
hep-th/9609076.} 

\lref\proyen{
B. de Wit, P. Lauwers and A. Van Proeyen, ``{\it Lagrangians of
N=2 Supergravity-Matter systems}'', Nucl. Phys. {\bf B255} (1983) 569.}

\lref\hawkingrad{
S. Hawking, ``{\it Particle Creation by Black Holes}'',
 Comunn. Math. Phys. {\bf 43} (1975) 199.}

\lref\complementarity{ G. 't Hooft, ``{\it The Black 
Hole Interpretation of String theory}'', Nucl.Phys {\bf B 335} (1990)
138;  G. 't Hooft, {\it The Scattering Matrix Approach for the 
Quantum Black Hole: An Overview}'' Int. J. Mod. Phys. {\bf A11}
(1996)4623, gr-qc/9607022;
 Y. Kiem, H. Verlinde
and E. Verlinde,
``{\it Black Hole Horizons and Complementarity}'',
 Phys.Rev.{\bf D52}, (1995) 7053 hep-th/9502074.}

\lref\boosts{D. Lowe, J. Polchinski, L. Susskind, 
L. Thorlacius and J. Uglum, 
``{\it  Black Hole Complementarity vs. Locality}'',
Phys.Rev. {\bf D52} (1995) 6997, 
 hep-th/9506138.}

\lref\intrilligator{K. Intrilligator and N. Seiberg, ``{\it Mirror Symmetry in 
Three Dimensional Gauge Theories}'', hep-th/9607207.}

\lref\alvarez{ L. Alvarez-Gaume and D. Freedman, ``{\it 
Geometrical Structure and Ultraviolet Finitness in the Supersymmetric
Sigma Model}'', Comm. Math. Phys. {\bf 80} (1981) 443, and references
therein.}

\lref\dkps{
M. Douglas, D. Kabat, P. Pouliot and S. Shenker, 
``{\it  D-branes and Short Distances in String Theory}'',
hep-th/9608024.}

\lref\bfss{
 T. Banks, W. Fischler, S.  Shenker and  L. Susskind,
``{\it  M Theory As A Matrix Model: A Conjecture}'', hep-th/9610043.} 

\lref\ghr{ S. Gates, C. Hull and M. Ro\v{c}ek, ``{\it
Twisted Multiplets and New Supersymmetric Non-Linear
$\sigma$-Models}'', Nucl. Phys. {\bf B248} (1984) 157.
}


%
\Title{\vbox{\baselineskip12pt
\hbox{hep-th/9611125}\hbox{RU-96-102}}}
{\vbox{
{\centerline { D-branes and  Near Extremal Black Holes }}
{\centerline {at Low Energies } }
  }}
\centerline{Juan Maldacena\foot{malda@physics.rutgers.edu}}
\vskip.1in
\centerline{\it Department of Physics and Astronomy, Rutgers University,
Piscataway, NJ 08855, USA}
\vskip.1in
\vskip .5in
\centerline{\bf Abstract}
It has been observed recently that  many properties of 
some near extremal black holes can be described in terms of 
 bound states of D-branes.
Using a non-renormalization theorem
 we argue that   the D-brane description
is the correct quantum gravity description of the black hole  at   
 low energies.
 The low energy theory  includes the black hole degrees
of freedom that account for the entropy and  describes also 
Hawking radiation.
The description is unitary and there seems to be no information loss
at low energies.

 \Date{}

%

\newsec{ Introduction }

Recently \ascv\ the entropy of extremal black holes in string theory was
calculated by counting the number bound states of D-branes.
The D-brane description corresponds to the weak coupling limit while
the black hole description corresponds to strong coupling. In the
first
case the gravitational radius of the configuration is smaller than the
string scale while it is bigger than the string scale for the latter.
Extremal black holes are supersymmetric BPS solutions. Supersymmetric
nonrenormalization arguments ensure that we can do the counting of 
states at small coupling and then extrapolate the result to the strong
coupling domain. This ensures that the D-brane counting agrees with
the classical area law for the black hole entropy \ascv .

While this explains the agreement found for extremal BPS solutions
\refs{\ascv , \spin , \sm , \clifford } it has not been clear why D-brane
calculations for near extremal black holes also agree with black
holes.
The agreement includes entropy counting \refs{ \ghas , \cm , \hlm ,
\becspin, \ktmbranes }
as well as more detailed dynamical properties such as absorption cross
sections and Hawking radiation \refs{  \tata , \dm , \gk ,
\jmas , \gkfour  , \cgkt}.

Here we give a rationale for this agreement for a class of 
near extremal five
dimensional black holes (in the so called dilute gas region).
The excitations of the D-brane system at low energies are described
in terms of a moduli space approximation. Using a non-renormalization
theorem we argue that this low
energy theory receives no corrections when we increase the coupling
and we go from the D-brane region into the black hole region. 
Therefore the same moduli space describes the 
low energy dynamics in the black hole region.
We then argue that the energy of the excitations accounting 
for the entropy and Hawking radiation are low enough to be described
within the low energy field theory. In order to do this we estimate
the
size of the corrections to the low energy theory, we estimate this
on the weakly coupled side and we see that extending this criterion
to the strong coupling region gives a sensible picture. 

We start in section 2 by describing the regime of interest, the 
type of black holes considered as well as the low energy condition.
In section 3 we describe the 
low energy D-brane theory and argue that it can be extrapolated to
strong coupling, provided the energy is low enough, we also give 
the condition that the energy has to satisfy. In section 4 
we explain why things calculated in the two regimes should agree.
In section 5 we study the possibility of D-brane emission. 
In section 6 we argue that these results imply that the dynamics
for these black holes 
is unitary at low energies.

\newsec{ Low energy field theory}

We start with  type IIB string theory
compactified
on $T^5= T^4\times S^1$.
We  consider five dimensional black holes (or six dimensional long
strings) parameterized by the four  classical parameters $r_0,r_n, r_1,
r_5$, the four parameters correspond to three charges and the mass.
The explicit solution is written in \hms\ and we  follow the
conventions there.
The charges correspond to
 a system  of $Q_5$ D-fivebranes wrapped on $T^5$, $Q_1$ D-onebranes
wrapped on $S_1$ and  momentum $P=n/R$ along $S^1$.

We  consider the dilute gas region defined by \jmas\
\eqn\dilute{
 r_0,r_n \ll r_1,r_5 
}
for reasons that will become clearer later. 
In most of the discussion we take the size of the $T^4$ to be small,
of order $V_4 \sim \alpha'^2$ and 
$S^1$  very long (we will discuss what changes if $S^1$ is small
later on) and we take $\alpha'=1$ (all lengths are measured in units
of 
$\sqrt{\alpha'}$). 
We also take $Q_1 \sim Q_5 \sim Q$, all these approximations are
done for simplicity and clarity in the argument and it is
straightforward
to extend the arguments for more general values of $V_4$ and $Q_1 \not
= Q_5$. Under these conditions $r_1 \sim r_5$.
%
The typical gravitational radius of the black hole is $r_g^2 =
{\rm max}\{r_1^2,r_5^2\} \sim  g
Q $.
 The gravitational radius is defined by the condition that
the redshift  between a static observer and the asymptotic
observer becomes of order one.

In all our discussion the coupling $g$ is   small $g \ll 1$ so that
closed string effects are small. However the effective open string
coupling is $gQ$ since it is like a large $N$ gauge theory\foot{The 
open string coupling is $g_{open} \sim  \sqrt{g}$.} 
($N=Q$).
 When the coupling  is weak $gQ \ll 1$ then we are in the
domain of validity of the D-brane perturbation theory. If
$gQ\gg 1$ we say that the coupling is strong and 
 we are in the semiclassical black hole domain.
 Note that this
definition of strong coupling is not the strong coupling region
$g \gg 1 $ which is  present in usual discussions on string dualities. 
Here we have strong coupling because of the large number $Q$ of  branes.

 We also  consider the low energy field 
theory,  the theory were the energies of all particles
satisfy 
\eqn\lowen{
\omega^2 r_g^2 \ll 1 ~~~~~~{\rm or }~~~~~~~ \omega^2 g Q \ll 1 
}
For example, in a scattering process the energies of the particles
measured at infinity satisfy \lowen .
In this limit the Compton wavelength of the particle is much 
bigger than the size of the
 black hole, so the black hole appears effectively as 
a pointlike system  from the point of view of the low energy 
theory on the bulk.
Note that energies are low with respect to $1/r_g$ which can itself be very
low, for an astronomical size black hole this energy is extremely low,
in particular much smaller than the string scale, the compactification scale
and other microscopic  scales in the problem. 

In the low energy  black hole region one can  do calculations using
the method of quantum fields on  a fixed classical background,
this is the semiclassical domain, it is the domain in
which Hawking radiation occurs, for near extremal  black holes 
the wavelength of Hawking radiation is much bigger than the gravitational
size of the black hole, $1/T_H \gg r_g$.
 If $1 \ll r_0^2+ r_n^2,r_1^2,r_5^2 $ these calculations do not   receive any
 $\alpha'$ corrections\foot{ If 
$r_0 \gg 1$  it is easy to see  that there is a smooth horizon, of size bigger
than $\alpha'$, however  we could also have a smooth horizon in the 
extremal limit, $ r_0 = 0$, as long as $r_n \gg 1$ (in other words, as long
as we have three large charges ($n,Q_1,Q_5$)).}. 
One can compute Hawking radiation in this way, absorption cross 
sections, etc.
The traditional semiclassical view \hawkingrad \hawkingunitarity\
is that in this case we can only have a thermal description of the
system, the emitted particles 
do not know about the microscopic state of the black
hole.

There are however things that we cannot do in this low energy
domain, we cannot measure the local geometry, since waves have
wavelengths much greater than the gravitational  radius, the
observer at infinity cannot measure the precise  shape of the
metric outside the horizon. His measuring rod is longer than the 
black hole. For him the black hole is as  a pointlike  system  that
can absorb energy and radiate it back thermally.
It should be noted however that  the absorption cross section
depends on some features of the geometry, so it is in some sense
a measure 
of the geometry, but not detailed enough  to sense the precise form
of the metric.

\vskip 1cm
\vbox{
{\centerline{\epsfxsize=5in \epsfbox{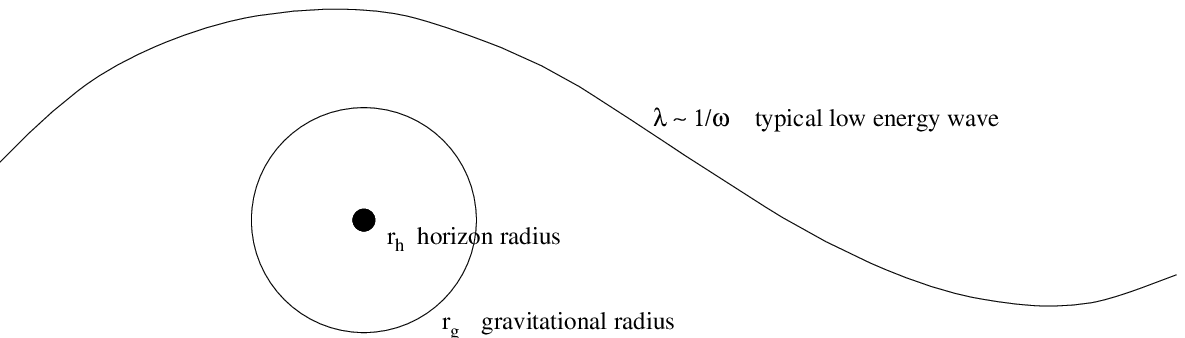}}}
{\centerline{\tenrm FIGURE 1: Various scales in the problem, with
$  r_0,r_n \ll r_1,r_5 \ll \lambda$. The sizes }}
{\centerline{ of the circles give an idea of the areas of the 
3-spheres and
$\lambda$ is the}}
{\centerline{ typical wavelength of the   particles we scatter.
}}
}
\vskip .5cm

\newsec{ D-brane low energy theory, Moduli space approximation.}

We will now concentrate on the open string sector of the theory
describing the excitations of the D-branes. This sector becomes
strongly coupled in the black hole region. 
This theory is a (1+1) dimensional field theory with (4,4)
supersymmetry
since this is the supersymmetry left unbroken by the extremal
D-branes (1D + 5D branes).  This supersymmetry is similar to N=2 in D=4.
These theories have vector multiplets and hypermultiplets.
In two dimensions the vector multiplet and hypermultiplet seem
very similar, both have four physical scalar components. 
The distinction between them is that they have  different transformation
properties under $R$ symmetries. This was discussed in 
the context of three dimensional theories in \seibergthree\ where the
same problem appeared. 
To understand this it is useful to think of this theory 
 as the dimensional reduction of a six dimensional
$N=1$
theory. In six dimensions there is a SU(2)$_R$ symmetry,
the vector multiplet has no scalar
components and its bosonic components are trivial under the SU(2)$_R$.
On the other hand the hypermultiplet has four scalar components
transforming as  the {\bf 2 } of SU(2)$_R$. When we reduce to
two dimensions we have an extra SO(4) $\sim$ SU(2)$_{\tilde L}\times$ 
SU(2)$_{\tilde R}$  
R-symmetry, again the vector and hypermultiplets will transform
differently under these R-symmetries and that is what distinguishes
them. It is interesting also that the two SU(2) factors coming from
SO(4) are correlated with the chirality in the 1+1 dimensional theory,
$\tilde L $ and $\tilde R $ denote also left and right moving. 
The vector multiplets are related to separation of the branes in
the extended four spatial dimensions and the hypermultiplets 
correspond to ``dissolving'' the onebranes inside the fivebrane
\douglas . This SO(4) symmetry of the gauge theory corresponds 
the SO(4) rotational symmetry of the five dimensional black hole \spin .

When we go to low energies we will keep only the massless
excitations and terms in the Lagrangian which are at most 
quadratic in the velocities. 
The D-brane low energy theory consists $4Q_1Q_5$ massless 
 fields parameterizing the moduli space of the bound
state of 1 branes and 5 branes. The moduli space is topologically
\eqn\modspace{
{\cal M } = (T^4)^{Q_1Q_5}/S(Q_1 Q_5)
}
where $S(m)$ is the permutation group of $m$ elements. This moduli 
space was obtained by duality arguments (by Vafa \vafainst ) and it was later
shown in \verlindecount \verlindemoore ~that this gives a microscopic 
counting of BPS  states with charges $n,Q_1,Q_5$ 
 which is fully U-duality
invariant. Summarizing, the situation is that we know by indirect
arguments
that the moduli space should be \modspace , at least topologically.
In principle one could calculate the metric on this moduli space 
in the weakly coupled D-brane theory.

As in four \proyen\ and three \intrilligator\ dimensions it
is possible to prove that supersymmetry implies that there
are no couplings between vectors and neutral hypermultiplets. 
A simple way to see this the following\foot{
I thank D. Kabat for pointing out an error in my previous argument.},
 first  we  choose
two left moving and
two rightmoving supercharges  out of the (4,4) available
and we realize explicitly  a (2,2) supersymmetry
by using (2,2) superfields.
In terms of (2,2) superfields the hypermultiplet decomposes into
a pair of chiral multiplets $\phi_h$ and the vector decomposes into a 
chiral multiplet $\phi_v$ and a twisted chiral multiplet $\chi$ \ghr .
The general (2,2) Lagrangian for chiral and twisted chiral 
fields was considered in \ghr . It is determined by a 
single function $K(\phi_p, \bar\phi_q, \chi_a, \bar\chi_b)$ which
gives  the metric and antisymmetric tensor field (B-field)  of
a non-linear sigma model
\eqn\metric{\eqalign{
G_{p\bar q} =&  \p_p \p_{\bar q} K~,~~~~~~~  
G_{a \bar b} =- \p_a \p_{\bar b} K \cr 
B_{p \bar b} =& \p_p \p_{\bar b} K~,~~~~~~~ 
B_{ \bar q a } = \p_{\bar q}\p_a K }}
 and the rest vanishes,
including the metric components mixing the chiral and twisted chiral
multiplets $G_{p \bar b} =0$ and $B_{p\bar q} = B_{a \bar b} =0$.
If we now perform a SU(2)$_{\tilde L}$ rotation on the system we can 
define new  (2',2') charges so that 
the 
chiral multiplets coming from the vector multiplet become
 twisted chiral and vice versa, $ \phi'_v =\chi ,~~ \chi'=\phi_v $,  
the chiral multiplets coming from
the hypermultiplets $\phi_h$ stay as chiral multiplets.
Combining the constraints of (2,2) invariance with (2',2') invariance 
we conclude that the metric and B-field components  mixing the
hypermultiplets
 with the vector multiplets
vanish, $G_{\phi_h  \phi_v} = 
G_{\phi_h \chi } = B_{\phi_h \phi_v} = B_{\phi_h  \chi } = 0$.
 Using eqn. \metric\ we see  that 
the sigma model  factorizes $K= K(\phi_h ,\bar \phi_h ) + 
  K(\phi_v ,\bar \phi_v , \chi ,\bar \chi ) $. The hypermultiplet
metric
is then  hyperk\"ahler since the sigma model  has (4,4)
supersymmetry and it has no torsion (B-field) \alvarez . 
The vector multiplet
moduli space corresponds to the models studied in \ghr\ and it is a 
generalized ``hyperk\"ahler'' manifold, which in some cases
can be reduced, via a duality transformation, to a usual 
hyperk\"ahler manifold \ghr . 
In any case, the conclusion is that the hypermultiplets are decoupled
from the vector multiplets\foot{
They are decoupled locally but there could be some gobal
identifications, which can usually be seen classically and
will not affect our later argument.
I thank N. Seiberg for pointing this out to me.}.

We are interested in the hypermultiplet moduli space since it
parameterizes
the space of possible bound state configurations \ascv \jmthesis .
Following the ideas in \seibergho\ we  regard the coupling
constant
as a background field, which should then be a vector multiplet
since it appears in front of the gauge kinetic term, an
interaction that would be forbidden if it were a hypermultiplet.
This implies that there are 
 no corrections, perturbative or non perturbative,
 to the hypermultiplet moduli space. 
This implies that the hyperk\"ahler  metric,
once we calculate
it,
is not renormalized when we increase the coupling.

In two dimensions we also have to worry about the fact that 
 vacuum expectation values are not well defined for
massless fields. 
It is more accurate to speak about the resulting
conformal field theory rather than the moduli space itself.
It is a conformal field theory because a hyperk\"ahler metric is
Ricci flat \alvarez . 
The statement would be  that the conformal field theory can be extrapolated
from weak to strong coupling. However there is another 
related  problem which  is that  the branches on the moduli 
space are not so well separated. 
There is a nonvanishing probability for the 
system to wander into the vector moduli space, which 
corresponds physically to the emission of D-branes, the scalars
of the vector multiplet correspond to separating the brane
in the extended $R^4$ spatial dimensions.  We will 
argue in sec. 6  that this process is highly suppressed for entropy 
reasons.  
Similar problems appear when non-renormalization theorems are
applied to the quantum mechanics of D0-branes \dkps \bfss .

As a aside, notice that there are indeed 
corrections to the vector moduli space,
for example if a one brane is far from the  fivebranes then the 
moduli space is classically flat but there is a one loop correction
coming from integrating out the massive (1,5) strings that gives
the $g Q/r^2$ correction to the metric in moduli space \metricfive .
This also shows that the coupling constant is indeed in a vector multiplet,
otherwise it could not have affected the vector multiplet moduli
space.

Note that the ``D-brane theory'' that has been applied to compute
the entropy \ghas , and scattering cross sections \dm , \jmas ,
 was precisely
this moduli space approximation to the motion of the D-branes since
only the massless excitations on the branes were taken into account. 
So it is   this moduli space  approximation 
that has been observed, by direct calculation,
to agree with  the semiclassical  results at strong coupling.

The conclusion is then that at low enough energies the 
excitations of the system are correctly described by this moduli 
space approximation, even for strong coupling!.
Now the question is: what energies are ``low enough''? 


First let us estimate, in the weak coupling theory, what the
mass of the least massive states is.
One appealing  picture is to think of the one brane charge as
carried by instantons on the fivebrane gauge theory \douglas .
However this parameterization is physically reasonable  only
when $Q_1 \ll Q_5$ (more precisely $r_1^2 \ll r_5^2$)
 otherwise the total energy in the instantons
in comparable to the energy of the fivebranes and the fivebrane
might bend or deform where there are many instantons. In 
other words, higher order terms in a Dirac-Born-Infeld-type  action
of the fivebrane  might  be important. 
In the case of $Q_1 \sim Q_5 $ it seems more reasonable to 
consider a set of two intersecting three branes (intersecting along
the $S^1$). Then the massless  degrees of freedom are somehow associated
to the $ Q_1 Q_5 $ intersection lines.
The transverse space of each set of three branes is a
two torus (say of size $\alpha'$). The three branes look like
points on this two torus.  If we assume that the  $Q$ three branes 
are uniformly distributed we find that the distance
between one and the nearest neighbor  is typically 
 $ r^2  \sim 1/Q $, so that $ m^2 \sim 1/Q $. 

Corrections due to the massive modes will go like
\eqn\massive{
 g~{ \omega^2 \over m^2 }
}
This implies that the corrections due to the lightest massive mode
are proportional to 
\eqn\lightmass{
  g Q \omega^2 \ll 1
}
which is small in the regime defined by \lowen .
There are also some other possibly light modes like D-strings going
between two different threebranes, which would have a mass
$m^2 \sim {1 \over gQ } $ and an interaction strength 
of order one, giving corrections proportional to \lightmass\ again. 
There are points in the moduli space where some states could become
light, for example if two threebranes come close to each other. 
This seems to  affect a small fraction of the hypermultiplets ($Q$ of
them vs. a total of $Q^2$) therefore  it will result in a small correction.
Since  there is a large number of massive states 
 there are  large $N(=Q)$ effects  going like 
\eqn\largenumber{
  g Q { \omega^2 \over \bar m ^2}
}
where $\bar m $ is the average mass and $gQ$ is the effective large
$N(=Q)$ coupling. But 
  $\bar m^2  \sim 1 $ since the typical distance between any
two threebranes is of the order of the compactification volume, so that 
we get \lowen\ again.  
Presumably all other effects we could imagine would also be
proportional to \lightmass .

In the case that the radius of the circle is small the low energy theory
corresponds to a 1+1 dimensional field theory whose target space is
 the moduli space \modspace\ but now on small circle. The fact that 
we divided out by the permutation group enables us to have twisted 
sectors in the low energy conformal field theory 
which correspond to long  multiply wound ``fractional'' strings 
\verlindemoore . 
These twisted sectors support excitations whose energy gap is
much smaller than $1/R$ ($R$ is the radius of $S_1$). The gap actually
becomes ${1\over RQ_1Q_5}$ which is much smaller than $T_L, T_R$ in the limit
of large charges \mss . 

\vskip 1cm
\vbox{
{\centerline{\epsfxsize=4in \epsfbox{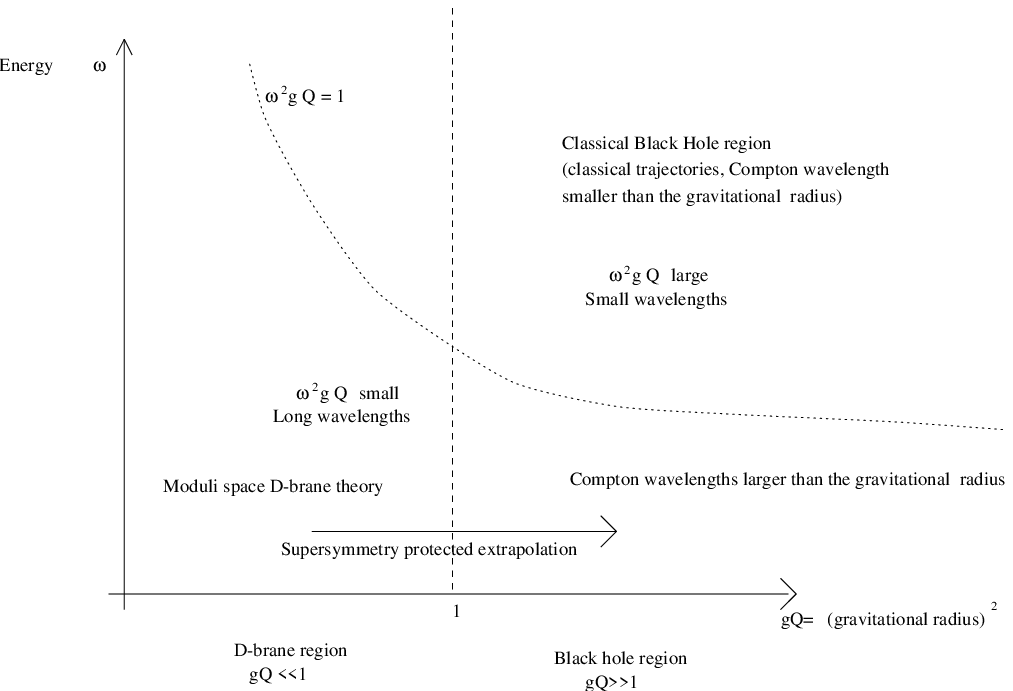}}}
{\centerline{\tenrm FIGURE 2: Different regions in the space of
parameters of  }}
{\centerline{a near extremal configuration. D-brane results can
be extrapolated for low energies. 
}}
}
\vskip .5cm

\newsec{D-brane vs. Black hole computations}

We saw in the previous section that the moduli space metric
for the system of one and fivebranes is not changed as 
we make the coupling strong. 
This non-renormalization theorem ensures that there are some
low energy processes that can be calculated in the strong coupling
regime (the black hole regime).
The entropy of the system will
be accurately given by the moduli space approximation if
the typical energy of the massless modes, which is proportional to
$ T_L,T_R$, satisfies  \lowen . This is indeed the 
case if $ r_0,r_n \ll r_1,r_5 $ since the temperatures are bounded by
 $  T_{L,R}  \leq  {\sqrt{r_0^2 + r_n^2} \over  r_1 r_5 }  \ll 1/r_g
$
\ghas .
So we conclude that the entropy is accurately given by the D-brane 
moduli space approximation, provided we are in the dilute gas region
\dilute . 

Now let us turn to  the scattering processes 
considered in \dm , \jmas .
The scalar considered there was an internal component of the metric $h_{ij}$
 of the
four torus. Since this metric appears in the moduli space metric of
the low energy D-brane theory,
we conclude that its coupling to the massless degrees of freedom is
not renormalized.

The 
calculations \cgkt ~that probe the higher
order terms in the Nambu action might also be understood by using this
line of argument. 
The moduli space \modspace\ seems to imply that the excitations of the
system are ``fractional'' strings, this is indeed true for BPS states
 \verlindemoore . It seems natural that  these strings should  
couple to the background metric  with the Nambu action. This deserves
a more careful analysis.

\newsec{Black hole fragmentation\foot{
Many of the remarks in this section originated in discussions with
A. Strominger.}.}

One of the possible decay modes of a black hole is by emission
of charged particles, by which the black hole loses its charge, in
some sense it fragments into the elementary constituents. 
In principle it can emit KK momentum, one brane winding charge and
fivebrane charge, the first one can be described in 
the D-brane  moduli space approximation described above \gk , \jmas\  and
 the last two correspond to some D-brane leaving 
the system. If we have a near extremal black hole which carries
three charges large  $n, Q_1, Q_5 $ then there is an entropic suppression
factor for the emission of charged particles. For example if it  emits
one unit of KK momentum, then the change in the extremal 
black hole entropy is 
$ \delta S = \pi \sqrt{ Q_1 Q_5 /n } $. The emission amplitude
therefore has a phase space 
suppression factor $e^{-\delta S }$.
If all charges are
large
then $\delta S $ is very large. This is independent of whether we are
in the dilute gas approximation or not, here the question is whether
the quantized values of the charges are large or not. 

In our discussion above we have ignored the possibility that
the D-branes leave the black hole since all our discussion
concentrated
on the hypermultiplet moduli space. 
As long as $R$ is not too small $R \ge \alpha'$ D-brane emission
will be suppressed because of the change in entropy  
$ \delta S \sim \pi \sqrt{ Q_5 n_{L,R} \over Q_1 } $ which 
is large in the region $ r_0^2 + r_n^2 \gg 1
$ corresponding to black holes with smooth horizons and small
$\alpha'$
corrections.
The conclusion is that the D-brane system corresponding to smooth, big, 
classical black hole solutions always have large $n_L$ or $n_R$ so that
D-brane emission is suppressed. Momentum emission (KK charge) could or 
could not be suppressed\foot{The situation is U-duality asymmetric
because
we are insisting in $g\ll 1$ and $R\ge \alpha'$.}.
Indeed, if the radius of $S_1$ is very large, then $n$ can be very large while
we are still in the dilute gas region. In this case KK charge emission is
not suppressed and the black hole, more properly a black string, will
discharge.
In the case that $R$ is small, say of the order of $\alpha'$ then 
we could take all $Q_1, Q_5 , n$ to be large in fixed proportions,
then
the charged emission will be suppressed. This can be intuitively
understood by 
remembering that all charged particles would have large masses so they
are not likely to be emitted. In fact, the suppression factor goes like
$e^{ -1\over R T_L}$ \jmas . It is very  important that 
in this case  (small $R$) the moduli 
space includes twisted sectors representing multiple windings, ensuring
a small energy gap and the existence of the low energy
excitations that account for the entropy. Notice that the total
energy of the  excitations on the moduli space is large  but
the temperature is small, due to the large number of degrees of
freedom. 

These reasons explaining why D-brane emission is suppressed 
 also justify our
restriction to the hypermultiplet moduli space in the D-brane
analysis of sec. 4.  

It seems that the best scenario for discussing the excitation and
decay of an extremal black hole is the one with $R$ small and large
$Q_1, Q_5 , N$, since in this case the black hole does not fragment,
it has a smooth geometry from the classical point of view and 
can be described by the D-brane moduli space as long as we 
are in the dilute gas and low energy regions \lowen . 

\newsec{Information Loss}

We have shown above that starting with a D-brane system we can go to
strong coupling and still continue having the same description
at low energies. 
It includes back reaction and it keeps track of the 
black hole microstates. The description is unitary, the unitarity
problem disappears when we use the full string theory.

At the same time we have the traditional semiclassical description
of the black hole. Since both descriptions
pertain to the same physical object they should somehow agree. 
The semiclassical results are recovered when we trace 
over the black hole microstates provided by the D-brane description.
It is important here that we are restricting to low energies \lowen ,
 at low energies the black hole already looks like a pointlike
system, so that replacing it by the D-brane moduli space theory just 
amounts to providing a description of the black hole states and their
interactions with the outside world. This effective low energy
theory  is similar  in spirit to the low energy description of
the scattering of massless fermions off a magnetic monopole 
(Callan Rubakov effect)  \polchinski , where one replaces the monopole
by a rotator sitting at the origin. It is clear that the 
low energy D-brane moduli space  Hamiltonian is unitary, 
massive modes provide just small corrections. 
A big difference between the two descriptions is that the
D-brane description keeps track of the black hole microstates.
Only after tracing them out we get the usual thermodynamic
description.


There is an interesting question: what exactly 
is the problem  about the usual 
information loss argument in this case?
The answer is not totally clear, it is an interesting problem.
Hawking's thermal matrix \hawkingunitarity\ 
relies on tracing over the modes that go into
the black hole, the D-brane picture suggests that 
one should think of these modes as part of the black
hole
excitations, so it is not reasonable to trace over them if one
is  keeping track of the changes in the black hole microstate as
the radiation is emitted.  There have been many suggestions
in the literature of things that could be wrong like 
some non-locality of string theory at high relative boosts \boosts ,
 the ideas of black hole complementarity \complementarity , etc..

Even though this   argument says that  there is no information
loss at low energies there could indeed be information loss at higher
energies since the D-brane moduli space description
is valid only at low energies. 
So the general question remains open but  there is 
a corner (low energies)  from which it seems  eliminated. 

It would be nice to extend these arguments to near extremal 
four dimensional black 
holes.

{\bf Acknoledgements}

It is a pleasure to thank V. Balasubramanian, 
T. Banks, M. Douglas, D. Kabat, D. Lowe,  S. Mathur,
J. Polchinski,  N. Seiberg, S. Shenker, A. Strominger and C. Vafa
for discussions, valuable coments and suggestions. 
This work was  supported in part by 
DOE grant
DE-FG02-96ER40559.

\listrefs

\bye

Explicit scattering calculations \jmas \gkfour\ suggest that the 
left and right moving modes are not on the horizon but 
localized at a physical distance $ l > r_0,r_n$ from the
horizon. Note that the distance could, in particular be much larger
than $\sqrt{\alpha'}$. It does not seem to be a phenomenon asociated to the 
string scale, the physics is regulated by the temperatures of the
left and right movers. This remark should however be taken with 
very carefully  since, as we mentioned above, it is not possible to talk
about the precise shape of the geometry at distances smaller than
$r_g$ when the energies obey \lowen .

It seems that  one can  definitely conclude that information 
is not lost in this kind of process: Low energy excitation
and decay of a black hole in the dilute gas region. 
An explicit way of doing this would be to take the extremal black
hole (with $n=0$), put it in a bath of particles with some 
effective temperature $T \ll 1/r_g$ then we remove the bath. 
Then we conclude that if we know the initial state of the bath and
the initial state of the black hole then 
there will be some correlations between the
outgoing Hawking radiation and the final state of the bath.
From the point of view of the D-brane theory this initial and final
states (configurations with zero KK momentum) have some residual
entropy, futhermore they are also singular 
configurations from the point of view of the classical theory, but they
are configurations with very small entropy which 
  cannot account for all the entropy that would
be generated by the formation and evaporation back to extremality 
of a macroscopic black hole with $r_0 \gg 1$. 
The description is analogous to the description that one has for
 a piece of charcoal
when it is heated and then cools down by radiating the extra energy.
In both cases it seems quite difficult to keep track of the precise 
quantum state as the system radiates and the correlations are
very subtle due to the large number of states.

\newsec{Conclusion}

We have argued that the  D-brane moduli space  results can be trusted
in the  black hole region in the low energy domain, providing the 
correct quantum gravity description of black holes, at low energies.
In this region the black hole looks like a pointlike system which
can be effectively described by the D-brane moduli space approximation.
The quantum gravity description is unitary in this low energy limit. 

It is important to emphasize that the description  claimed here  is only
between the low energy limits of both systems.
The physics is however very rich in this limit since the degrees of freedom
which account for the  
black hole entropy and Hawking radiation are included  within the low energy 
approximation.

One of the most interesting questions is to understand better the 
relation between classical geometry and the thermal  D-brane moduli space
description. In this approach the classical geometry is describing the
thermodynamic approximation  of the system. Understanding this
more precisely seems important to pinpoint what is being overlooked 
in the semiclassical approximation. The approach taken here is 
indirect, providing an argument explaning why the D-brane description
is the correct quantum gravity description of the system  at low energies,
address this question. 

It would also be nice to extend these arguments to four dimensional
black holes. 

It seems that  one can  definitely conclude that information 
is not lost in this kind of process: Low energy excitation
and decay of a black hole in the dilute gas region. 
An explicit way of doing this would be to take the extremal black
hole (with $n=0$), put it in a bath of particles with some 
effective temperature $T \ll 1/r_g$ then we remove the bath. 
Then we conclude that if we know the initial state of the bath and
the initial state of the black hole then 
there will be some correlations between the
outgoing Hawking radiation and the final state of the bath.
From the point of view of the D-brane theory this initial and final
states (configurations with zero KK momentum) have some residual
entropy, futhermore they are also singular 
configurations from the point of view of the classical theory, but they
are configurations with very small entropy which 
  cannot account for all the entropy that would
be generated by the formation and evaporation back to extremality 
of a macroscopic black hole with $r_0 \gg 1$. 
The description is analogous to the description that one has for
 a piece of charcoal
when it is heated and then cools down by radiating the extra energy.
In both cases it seems quite difficult to keep track of the precise 
quantum state as the system radiates and the correlations are
very subtle due to the large number of states.

There is an interesting question: what is the problem  about the usual 
information loss argument in this case?
The answer is not totally clear, it is an interesting open problem.
Hawking's thermal matrix \hawkingunitarity\ 
relies on tracing over the modes that go into
the black hole, the D-brane picture suggests that 
one should think of these modes as part of the black
hole
excitations, so it is not reasonable to trace over them if one
us keeping track of the changes in the black hole microstate as
the radiation is emitted.

Explicit scattering calculations \jmas \gkfour\ suggest that the 
left and right moving modes are not on the horizon but 
localized at a physical distance $ l > r_0,r_n$ from the
horizon. Note that the distance could, in particular be much larger
than $\sqrt{\alpha'}$. It does not seem to be a phenomenon asociated to the 
string scale, the physics is regulated by the temperatures of the
left and right movers. This remark should however be taken with 
very carefully  since, as we mentioned above, it is not possible to talk
about the precise shape of the geometry at distances smaller than
$r_g$ when the energies obey \lowen .

Even though this  indirect argument says that  there is no information
loss at low energies there could indeed be information loss at higher
energies, so the general question remains open but it there is 
a corner from which it seems  eliminated. 

It would be nice to extend these arguments to near extremal 
four dimensional black 
holes.

{\bf Acknoledgements}

It is a pleasure to thank T. Banks, M. Douglas, D. Lowe,  S. Mathur,
J. Polchinski,  N. Seiberg, S. Shenker, A. Strominger and C. Vafa
for discussions, valuable coments and suggestions. 
This work was  supported in part by 
DOE grant
DE-FG02-96ER40559.

\listrefs

\bye